\documentclass[journal=jpca, manuscript=article]{achemso}
\usepackage[utf8]{inputenc}
\usepackage{dcolumn}
\usepackage[version=3]{mhchem}
\usepackage{paralist}
\usepackage{xcolor}
\usepackage{cancel}

\newcolumntype{d}[1]{D{.}{.}{#1}}
\newcommand{\ehomo}{\epsilon_\mathrm{H}}
\newcommand{\elumo}{\epsilon_\mathrm{L}}
\newcommand{\oip}{\omega_\mathrm{H}}

\newcommand{\ogdd}{\omega_\text{GDD}}

\newcommand{\jip}{J^2_\mathrm{H}}
\newcommand{\gddpbe}{PBE($\ogdd$)}
\newcommand{\gddpbeh}{PBEh($\ogdd$)}
\newcommand{\ippbe}{PBE($\oip$)}
\newcommand{\ippbeh}{PBEh($\oip$)}
\newcommand{\herbert}{PBEh($0.20$)}
\newcommand{\pbestandard}{PBE($0.40$)}
\newcommand{\rhos}{\rho_{\sigma}}

\newcommand{\taus}{\tau_\sigma}

\newcommand{\hxs}{h_\mathrm{X}^\sigma}
\newcommand{\hxssr}{h_\mathrm{X,SR}^\sigma}
\newcommand{\hxslr}{h_\mathrm{X,LR}^\sigma}

\newcommand{\exlr}{E_\mathrm{X,LR}}

\newcommand{\diff}{\mathrm{d}}

\newcommand{\rr}{\mathbf{r}}
\newcommand{\ra}{\mathbf{r}_1}
\newcommand{\rb}{\mathbf{r}_2}
\newcommand{\rab}{r_{12}}

\newcommand{\erfc}{\mathrm{erfc}}
\newcommand{\erf}{\mathrm{erf}}
\newcommand{\eqn}[1]{Eq.~\ref{#1}}

\newcommand{\dxs}{\langle d^2_\mathrm{X,\sigma} \rangle}
\newcommand{\dxsvec}{\mathbf{d}_\mathrm{X,\sigma}}

\newcommand{\citenumwrap}[1]{\nocite{#1}\citenum{#1}}
\newcommand{\citetwrap}[1]{\nocite{#1}\citet{#1}}

\author{Marcin Modrzejewski}
\affiliation[University of Warsaw]{Faculty of Chemistry, University of Warsaw, 02-093 Warsaw, Pasteura 1, Poland}
\email{m.m.modrzejewski@gmail.com}
\author{Łukasz Rajchel}
\affiliation[University of Duisburg-Essen]{Faculty of Chemistry, University of Duisburg-Essen, Universitätsstraße 5, 45117 Essen, Germany}
\alsoaffiliation[ICM, University of Warsaw]{Interdisciplinary Centre for Mathematical and Computational Modelling, University of Warsaw, ul. Prosta 69, 00-838 Warsaw, Poland}
\author{Grzegorz Chałasiński}
\affiliation[University of Warsaw]{Faculty of Chemistry, University of Warsaw, 02-093 Warsaw, Pasteura 1, Poland}
\author{Małgorzata M. Szczęśniak}
\affiliation[Oakland University]{Department of Chemistry, Oakland University, Rochester,
  Michigan 48309-4477, USA}

\title{Density-Dependent Onset of the Long-Range Exchange: A Key to Donor-Acceptor Properties}
\begin{document}
\begin{abstract}
Quantum mechanical methods based on the density functional theory (DFT) offer a realistic possibility of first-principles design of organic donor-acceptor systems and engineered band-gap materials.  This promise is contingent upon the ability of DFT to predict one-particle states accurately.  Unfortunately, approximate functionals fail to align the orbital energies with ionization potentials.  We describe a new paradigm for achieving this alignment. In the proposed model, an average electron---exchange hole separation controls the onset of the orbital-dependent exchange in approximate range-separated functionals. The correct description of one-particle states is thus achieved without explicit electron removal or attachment. Extensive numerical tests show that the proposed method provides physically sound orbital gaps and leads to excellent predictions of charge-transfer excitations and other properties critically depending on the tail of the electron density.
\end{abstract}


\section{Introduction}
A nonlocal exchange functional is critical to the description of properties that depend on the tail of the electron density, such as charge-transfer (CT) and Rydberg excitations,\cite{dreuw2003long,dreuw2004failure,tawada2004long} repulsive part of the van der Waals potential,\cite{kamiya2002density} and frontier orbitals.\cite{stein2010fundamental} In range-separated (RS) functionals,\cite{gill1996family,leininger1997combining,iikura2001long} a nonlocal, orbital-dependent exchange functional is introduced in the long-range part of the electron-electron interaction. The critical issue is at what distance the long range should commence. This paper presents a model that explains the underlying physics of a molecular system behind the proper onset of the long-range exchange. An efficient and reliable functional is derived for the properties depending on the density tail.

A mixture of a local density functional approximation (DFA) and orbital-dependent Hartree-Fock (HF) exchange was first proposed by Becke\cite{becke1993new} and subsequently incorporated into a variety of models. The non-local, orbital-dependent exchange absorbs a part of the missing discontinuity of the exchange-correlation potential\cite{perdew1982density} plaguing local DFAs.\cite{seidl1996generalized,yang2012derivative} Although hybrid DFAs improve on the quality of molecular properties, several issues remain. Most notably, qualitative errors in donor-acceptor properties, fundamental gaps from orbital energies, and CT excitations were attributed to the persistence of the many-electron self-interaction error.\cite{mori2006many,cohen2012challenges} These deficiencies of hybrid DFAs can be alleviated by applying different models for short- and long-range electronic interaction according to the decomposition of the electron-electron interaction operator:\cite{gill1996family,leininger1997combining,iikura2001long}
\begin{equation}
\frac{1}{r_{12}} = \frac{\erfc(\omega r_{12})}{r_{12}} + \frac{\erf(\omega r_{12})}{r_{12}}. \label{interaction-splitting}
\end{equation}
The decomposition is followed by the separation of short- and long-range parts out of the total exchange hole,
\begin{equation}
\hxs(\ra, \rb) = \hxssr(\ra, \rb) + \hxslr(\ra, \rb),
\label{hsx_split}
\end{equation}
where
\begin{equation}
\hxslr(\ra, \rb) = - \frac{\left|\sum_i^{N_\sigma}\psi_{i\sigma}^*(\ra) \psi_{i\sigma}(\rb)\right|^2}{\rho_\sigma(\ra)} \erf(\omega \rab).
\end{equation}
The partitioning in \eqn{hsx_split} is then mirrored in the exchange energy, e.g. its long-range part reads
\begin{equation}
  \exlr = \sum_\sigma \frac 1 2 \int \frac{\rho_\sigma(\ra) \hxslr(\rb) }{\rab} \diff \ra \diff \rb.
\end{equation}


Regardless of the specific value of $\omega$, modeling $\exlr$ entirely by the orbital-dependent HF exchange constrains the asymptote of an exchange potential to the Coulombic decay, as in the exact theory,\cite{baer2010tuned}
\begin{equation}
  \hat{K}_\mathrm{X} \psi_i(\rr) \xrightarrow{r \to \infty} - \frac{1}{r} \psi_i(\rr). \label{exact-asymptotic}
\end{equation}
where $\psi_i(\rr)$ denotes an occupied orbital and $\hat{K}_\mathrm{X}$ is the exchange operator.

A system-specific $\omega$ can be obtained non-empirically by means of optimal
 tuning.\cite{kronik2012excitation} For instance, $\omega$ can be adjusted so that
 the frontier (HOMO and LUMO) orbital energies, $\ehomo$ and $\elumo$, match as
 closely as possible ionization potentials (IPs) and electron affinities (EAs),
 respectively.\cite{kronik2012excitation,stein2009prediction}
Another variant of this method, IP-tuning, constrains only the HOMO energy:\cite{koerzdoerfer2011long}
\begin{equation}
  \jip(\oip) = \min_\omega \Big( E(N-1) - E(N) + \ehomo \Big)^2. \label{ip-opt}
\end{equation}

All quantities on the right-hand side of \eqn{ip-opt} depend on $\omega$. $E(N)$ denotes
 the total energy of an $N$-electron system. Other choices of the objective function may also
 also be justified and are sometimes recommended.\cite{refaely2012quasiparticle} \eqn{ip-opt}
 imposes Koopmans' theorem which is satisfied in the exact Kohn-Sham theory. 
Several authors demonstrated that tuning of an RS functional is a route towards reliable 
charge-transfer (CT) excitations,\cite{stein2009reliable,stein2009prediction,kronik2012excitation} 
HOMO-LUMO (HL) gaps,\cite{stein2010fundamental,kronik2012excitation} photoelectron
 spectra,\cite{refaely2012quasiparticle} and hyperpolarizabilities.\cite{sun2013influence} A
 survey of the solutions of \eqn{ip-opt} for organic
 polymers\cite{koerzdoerfer2011long,koerzdoerfer2012relationship} reveals that {$\oip$} dramatically
 depends on the system size and electronic structure.

The tuning procedure involves multiple evaluation of IPs, a step required to properly align
 single-particle spectra. In what follows, we will develop a model which enables a single-step
 computation of an approximation to the optimal RS parameter for a given system.

\section{Theory}
The physical picture behind the $-1/r$ asymptote of the exchange potential
  is an electron in the outer regions of a molecule
 interacting with its exchange hole, which is a compact charge distribution
 residing away from the reference electron, in the region of localized orbitals.
 (See Fig.~10 in Ref.~\citenumwrap{neumann1996exchange} for an illustration
 of the exchange hole interacting with an electron beyond the last occupied shell.)
 Such an interaction is properly accounted for by the HF exchange hole. Consequently,
 the average separation of the outer-density electron from its exchange hole
 is an estimate of the interelectronic distance at which the long-range part of 
 \eqn{interaction-splitting} should prevail. The average electron---exchange
 hole distance thus constitutes an upper bound on the value of $1/\omega$, which in turn is
 a measure of the distance at which the long-range part of~\eqn{interaction-splitting} becomes dominant. To
 proceed further, we assume that this condition not only bounds~$1/\omega$ from above,
 but completely defines the distance at which the transition between short- and
 long-range forms of the exchange hole occurs in an optimal RS functional.

We define the average squared electron---exchange hole distance as
\begin{equation}
\dxs = \frac{\displaystyle \int \rhos(\rr) w_{\sigma}(\rr) \dxsvec^2(\rr) \, \diff \rr} %
     {\displaystyle \int \rhos(\rr) w_{\sigma}(\rr) \, \diff \rr}, \label{dxs-def}
\end{equation}
where $\dxsvec$ is the vector pointing from the position of the reference electron
to the charge center of the HF exchange hole, $\rhos$ is the $\sigma$-spin electron
density, and $w_{\sigma}$ is a weight function. The weight function is required so that the density-tail region influences $\dxs$ more than the bulk region. 
The choice of~$w_\sigma$ is not uniqe. For instance, the formulas developed
for switching between bulk and asymptotic exchange-correlation potentials in the
context of asymptotic correction
 methods\cite{tozer1998improving,gruning2001shape} could be used. 
We decided to use a step function,
\begin{equation}
w_{\sigma}(\rr) = 
\begin{cases}
  1, & t_\sigma(\rr) \le \mu, \\
  0, & t_\sigma(\rr) > \mu,
\end{cases} \label{stepfunction}
\end{equation}
which is based on a simplified variant of the electron localization
 function\cite{becke1990simple} proposed by \citet{schmider2000chemical},
\begin{align}
  t_\sigma(\rr) &= \tau_{\mathrm{UEG},\sigma}(\rr)/\taus(\rr), \\
  \tau_{\mathrm{UEG},\sigma} &= \frac{3}{5}\left( 6 \pi^2 \right)^{2/3} \rhos^{5/3}, \\
  \taus &= \sum_i^{N_\sigma} |\nabla\psi_{i\sigma}|^2.
\end{align}
$t_\sigma$ is greater than $1$ in regions where localized orbitals dominate, and tends to zero in the
 density tail.\cite{schmider2000chemical} Thus, for a sufficiently small $\mu$, $w_\sigma = 1$ 
beyond any localized orbital shell, and $w_\sigma = 0$ elsewhere. Our preliminary results have shown that
for most systems satisfactory results are obtained if the value of $\mu$ is so chosen that $N_\sigma = 1$
 in the normalization integral,
\begin{equation} 
  N_{\sigma} = \int \rhos(\rr) w_{\sigma}(\rr; \mu) \diff \rr. \label{mu-param}
\end{equation}
The inequality $\mu \ge 0.07$ is enforced to prevent $\mu$ from attaining extremely small numerical values.
 To retain the size consistency of $\dxs$, $N_\sigma$ should be set to the number of noncovalently bound subsystems when $\dxs$ is 
computed for a noncovalent complex. The issue of the size consistency of system-specific RS functionals is discussed further in the text.
 The vector $\dxsvec$ in~\eqn{dxs-def} points to the center
 of charge of the exact, orbital-dependent exchange hole,~$\hxs$. It can be computed analytically at a cost
 no higher than any single-electron integrand on a molecular grid,\cite{becke2005exchange}
\begin{align}
\dxsvec(\rr) &= -\int \hxs(\rr, \rr') \rr' \diff \rr' - \rr \nonumber \\
&= \left( \frac{1}{\rhos(\rr)} \sum_{ij} f_{ij\sigma} \psi_{i\sigma}(\rr) \psi_{j\sigma}(\rr) \right) - \rr, \label{dxsvec-def} \\
f_{ij\sigma} &= \int \rr' \psi_{i\sigma}(\rr') \psi_{j\sigma}(\rr') \diff \rr'. 
\end{align}
Using \eqn{dxs-def} we define the global-density-dependent (GDD) RS parameter,
\begin{equation}
\ogdd = \frac{C}{\sqrt{\dxs}}, \label{ogdd-def} 
\end{equation}
where $C$ is a numerical constant depending on the underlying functional, the fraction
 of short-range HF exchange, and the weight function used in the definition of {$\dxs$}.
 It does not depend on a physical system. The parameter $C$ for any particular functional
 was determined via least-squares fitting of the model RS parameter, $\ogdd$, to the IP-tuned $\oip$.
 The fitting was performed within the set of molecules presented in Fig.~\ref{ip-figure-pbe}, excluding
 only the outliers: the noble gases, \ce{Cl2}, \ce{ClF}, and \ce{F2}. No empirical or
 higher level theoretical data were involved in determing $C$. The spin label is dropped from
 the left-hand side of \eqn{ogdd-def} because all systems studied in this work are closed-shell.
 In general case, however, $\ogdd$ would be spin-dependent.

 By PBE($\omega$) we denote an RS functional composed of PBE correlation,\cite{perdew1996generalized}
 the short-range PBE exchange by~\citetwrap{henderson2008generalized}, and the long-range HF exchange.
 PBE($\omega$) contains no short-range HF exchange. The definition of PBEh($\omega$) is the same as PBE($\omega$), 
 except for $20\%$ of the short-range HF exchange. {\herbert}, also known as LRC-$\omega$PBEh, is the functional
 developed by~\citet{rohrdanz2009long}. In our experience, PBE($0.40$) performs almost identically to LC-$\omega$PBE
 proposed by~\citetwrap{vydrov2006assessment}, although it differs in the derivation of the short-range exchange.
 The $C$ parameters used to compute the GDD-based RS parameters for {\gddpbe} and {\gddpbeh} are $0.90$ and
 $0.75$, respectively. 
 
 The value of $\ogdd[\rho]$ need not be
 evaluated fully self-consistently. We recommend the following protocol:
 \begin{inparaenum}[(i)] \item Perform a single point calculation using {\pbestandard} or {\herbert}.
 \item Use the converged density to compute $\ogdd[\rho]$ and provide a guess for the {\gddpbe} or
   {\gddpbeh} calculation. \end{inparaenum} A single cycle of this procedure is sufficient for well-converged
 results. All numerical results presented in this work were generated in this way, except for those
 in Figs.~\ref{ip-figure-pbe} and~\ref{alkanes-figure}, where densities from self-consistent IP-optimized
 calculations were used. All calculations employed def2-TZVPP
 basis\cite{weigend2005balanced,schuchardt2007basis} unless specified otherwise.

\section{Results and discussion}
\begin{figure}[H]
\includegraphics[width=0.5\textwidth]{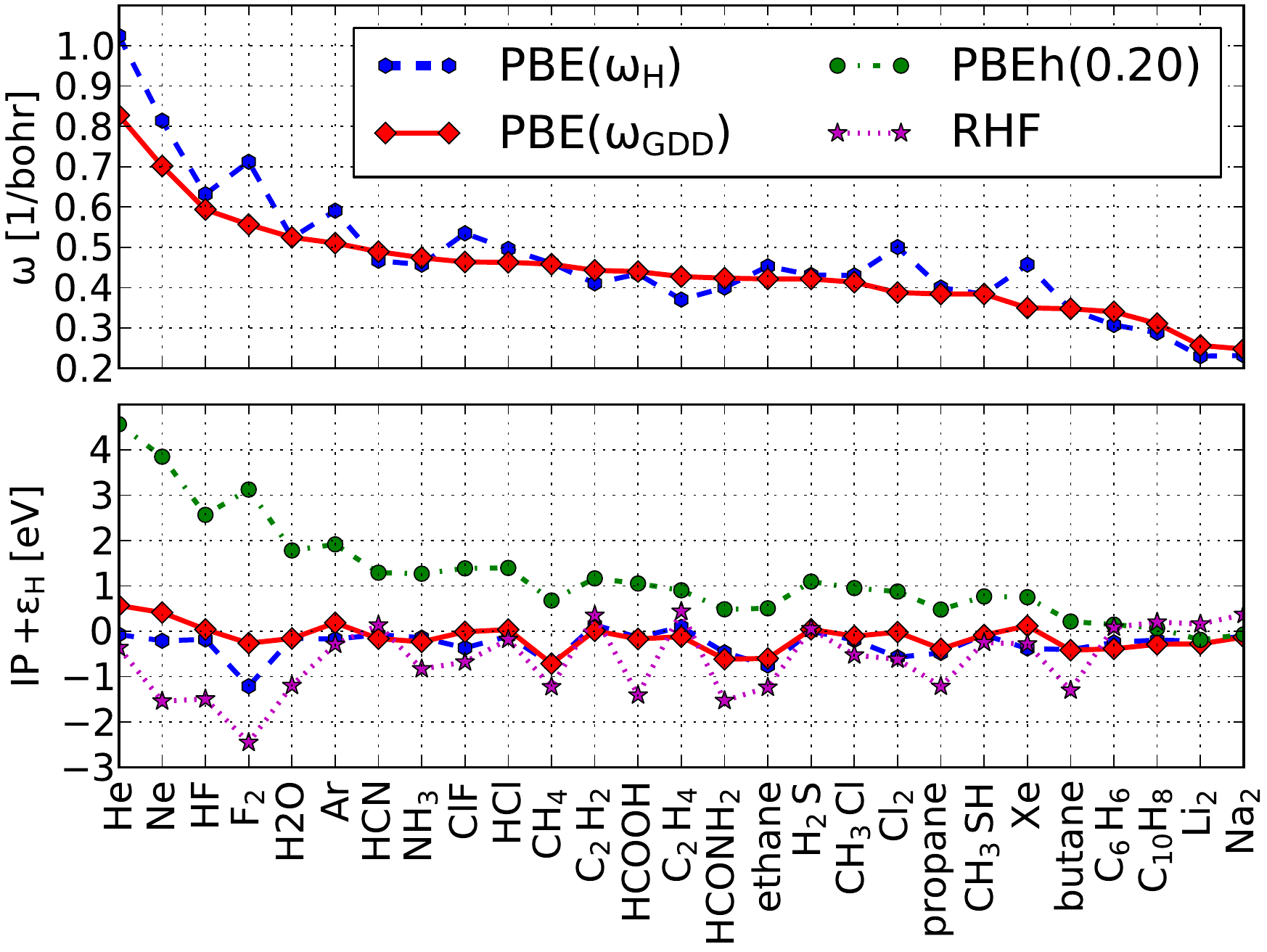}
\caption{Comparison of IP-optimized and GDD-based RS parameters for the PBE($\omega$) functional (upper panel). Deviation of HOMO energies from experimental IPs\cite{linstrom2013nist} (lower panel). \ce{C10H8} denotes naphthalene.}\label{ip-figure-pbe}
\end{figure}

\begin{figure}[H]
\includegraphics[width=0.5\textwidth]{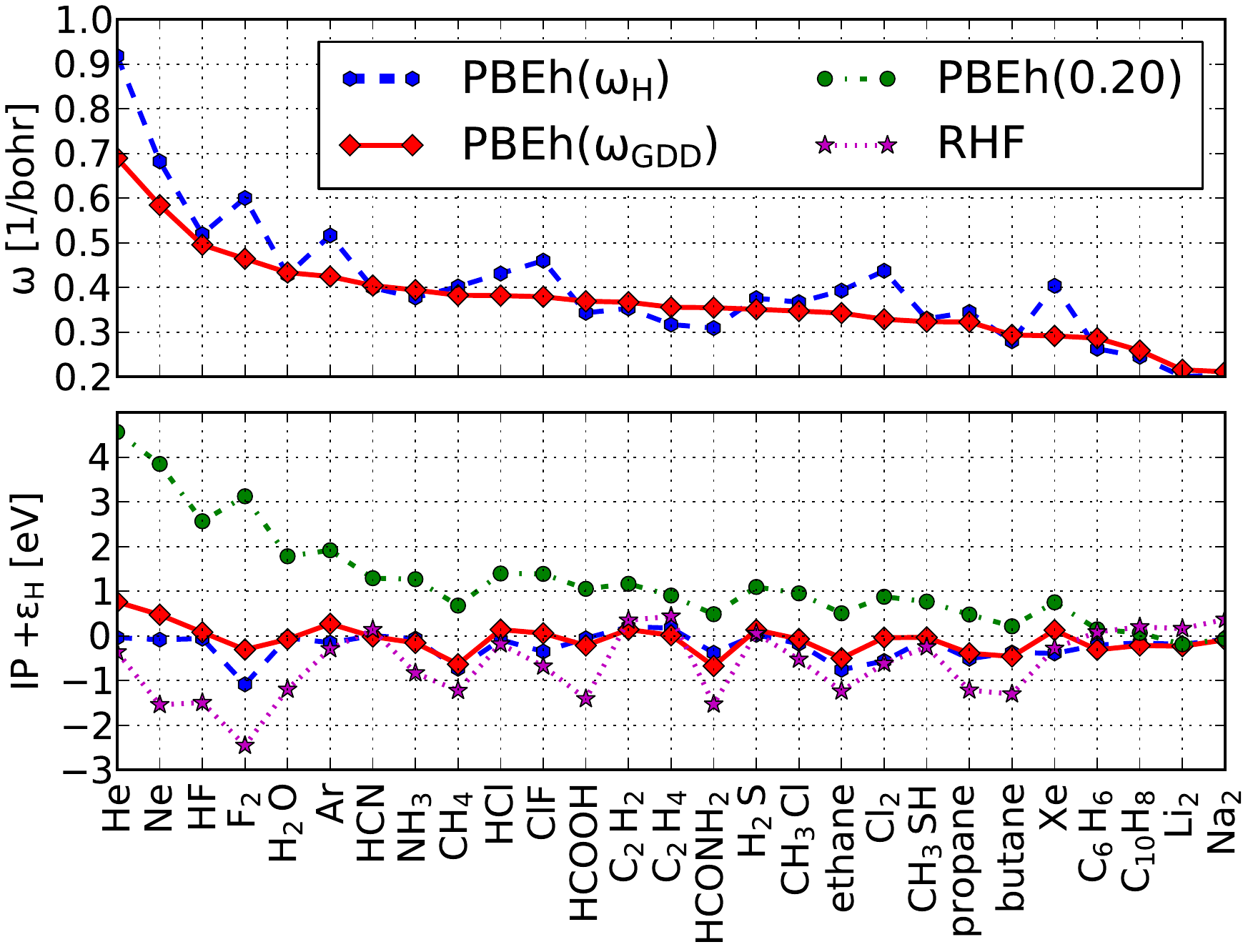}
\caption{Comparison of IP-optimized and GDD-based RS parameters for the PBEh($\omega$) functional (upper panel). Deviation of HOMO energies from experimental IPs\cite{linstrom2013nist} (lower panel).}\label{ip-figure-pbeh}
\end{figure}

\citetwrap{rafaely2011fundamental} observed that the optimal RS parameter should decrease as the characteristic radius of a molecule grows. This trend is captured by $\ogdd$, which mimics the behavior of $\oip$ across the wide range of systems, from single atoms to multiple aromatic rings, see Fig.~\ref{ip-figure-pbe}. The differences between $\oip$ and $\ogdd$ are small, except for halogen diatomics: \ce{F2}, \ce{Cl2}, and \ce{ClF}. For these systems $\ogdd$ deviates substantially from $\oip$, but it is {\gddpbe} which leads to a better agreement of $-\ehomo$ with the experimental IPs. The advantage of {\gddpbe} stems from the fact that IPs are obtained directly from Koopmans' theorem, without involving any energy differences of neutrals and ions. This is important, since for the halogen diatomics these energy differences are of poor quality. Overall, both {\ippbe} and {\gddpbe} yield small mean-absolute-percentage deviations (MAPDs) of $\ehomo$ energies from experimental IPs, $2.5\%$ and $2.1\%$, respectively. The signed errors of {\gddpbe} and {\ippbe} are bracketed by two curves: one corresponding to the HF method, and the other to the {\herbert} functional of~\citetwrap{rohrdanz2009long}. Although the characteristic radius of a molecule is the decisive factor in most cases, there exists a special case of the alkali diatomics, \ce{Li2} and \ce{Na2}, for which both GDD and IP-optimized RS parameters reach the smallest values in our set of molecules, even smaller than in a more extended naphthalene. Although the RS parameters of {\gddpbeh} and {\gddpbe} differ due to the varying amount of short-range HF exchange, the performance of these methods with respect to Koopmans' theorem is almost identical, see Figs.~\ref{ip-figure-pbeh} and~\ref{ip-figure-pbe}.

Having established that the GDD-based functionals yield HOMO energies close to the IPs, we now turn to the interpretation of LUMO energies. In practical applications, the total energy in an RS approximation is minimized with respect to orbitals and not with respect to the electron density, unlike traditional Kohn-Sham theory.\cite{kohn1965self} Conseqently, the asymptote of the exchange operator depends on whether $\hat{K}_\mathrm{X}$ acts on an occupied or virtual orbital. This is similar to the well-known property of the exchange operator in HF theory. Formally, such a functional is defined within generalized Kohn-Sham (GKS) theory.\cite{seidl1996generalized} If we consider an electron in an $N$-electron system which is far from the nuclei and other electrons, at the distance~$r \rightarrow \infty$, it feels the attraction from the nuclei and the repulsion from the remaining~$N - 1$ electrons. In the RS DFT model, the electronic potential the electron feels is~$(N - 1)/r$, where $N/r$ is contributed by the Coulomb operator and $-1/r$ by the exchange operator to compensate for the Coulombic self-interaction. If an electron is attached to the $N$-electron system, it feels the repulsion of the remaining $N$ electrons. Such a picture is properly captured in the GKS with the extra electron occupying LUMO: the Coulomb operator now contributes no self-interaction, therefore the exchange potential does not decay as $-1/r$ when acting on a virtual orbital. The interpretation of LUMO as an orbital of the attached electron is supported in GKS and is not supported by the traditional Kohn-Sham theory, where the exact exchange potential is multiplicative and always decays as $-1/r$.

\begin{figure}
  \includegraphics[width=0.5\textwidth]{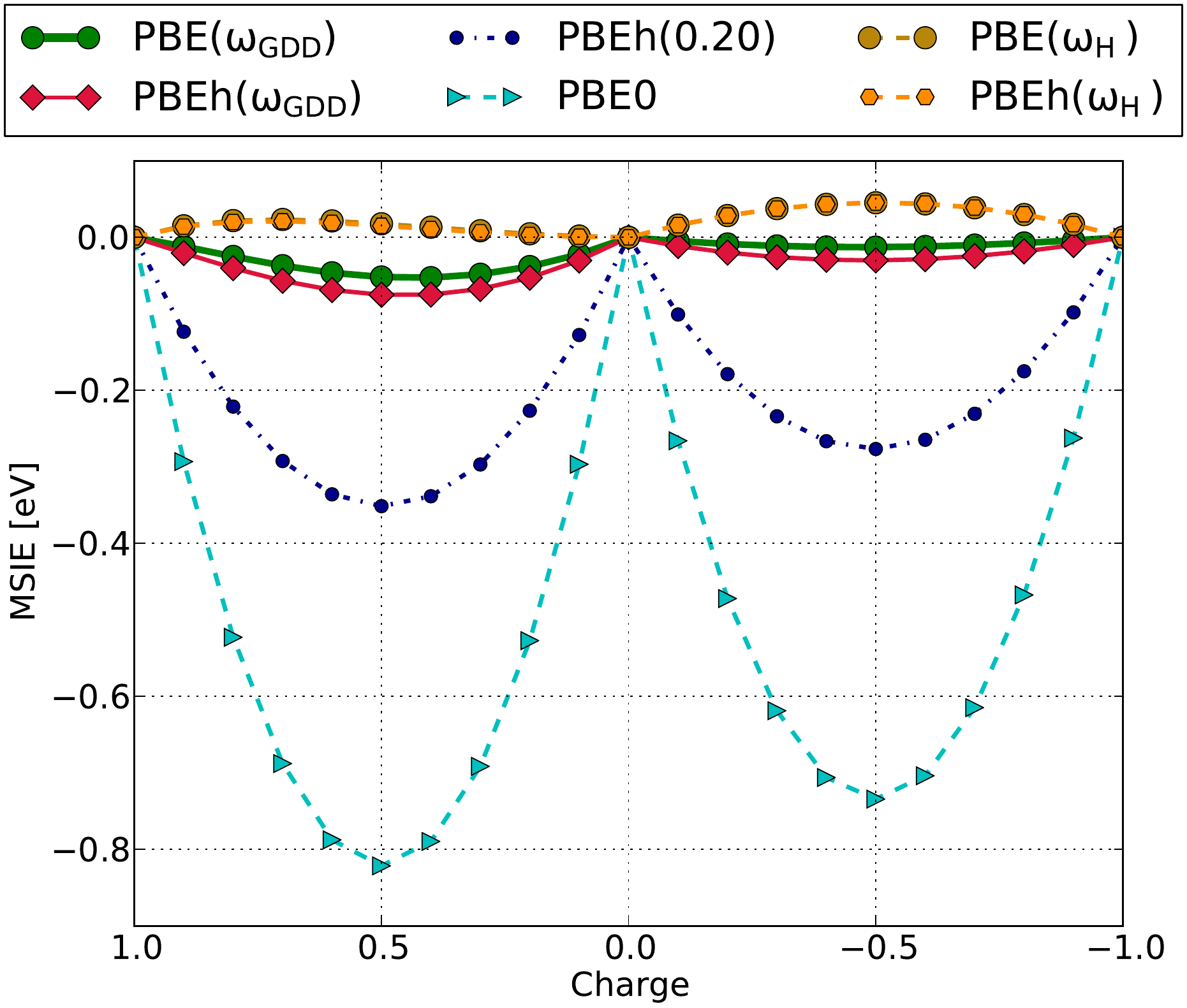}
  \caption{Deviation of $E(N)$ from straight line for \ce{ClF} molecule.} \label{msie-clf}
\end{figure}

\begin{figure}
  \includegraphics[width=0.5\textwidth]{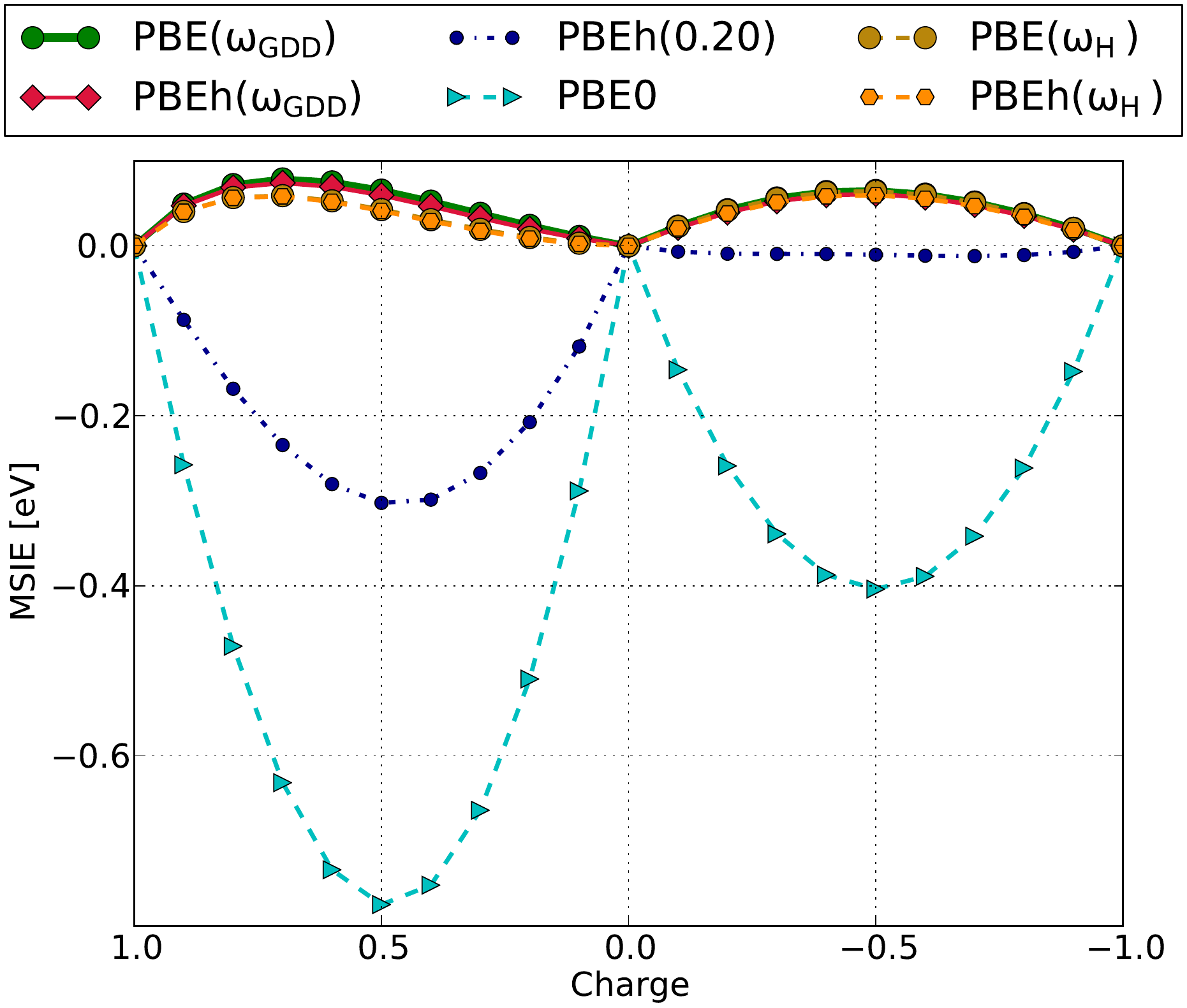}
  \caption{Deviation of $E(N)$ from straight line for \ce{NH3} molecule.}\label{msie-nh3}
\end{figure}

It was proven by~\citet{yang2012derivative} that for RS functionals within GKS framework, $\elumo$ is the slope of $E(N)$ between a neutral and an anion,
\begin{equation}
  \left( \frac{\partial E}{\partial N} \right)_\upsilon^{(+)} = \elumo.
\end{equation}
Thus, if $E(N)$ is a piecewise-linear curve, $-\elumo$ can serve as an approximation to the EA. A deviation from the straight line is called the many-electron self-interaction error\cite{mori2006many} (MSIE). Figs.~\ref{msie-clf} and~\ref{msie-nh3} show that the MSIE of a conventional hybrid functinal (PBE0\cite{adamo1999toward}) is reduced several times by introducing RS exchange in the fixed-$\omega$ {\herbert} functional. Note that \ce{ClF} has a positive EA, whereas \ce{NH3-} is bound only by the finite basis set. Further reduction of the error occurs when we use a system-dependent RS parameter instead of a fixed value. Therefore, both IP-tuned and GDD-based functionals are characterized by an order-of-magniutude lower MSIE than the PBE0 functional.

\begin{figure}[H]
\includegraphics[width=0.5\textwidth]{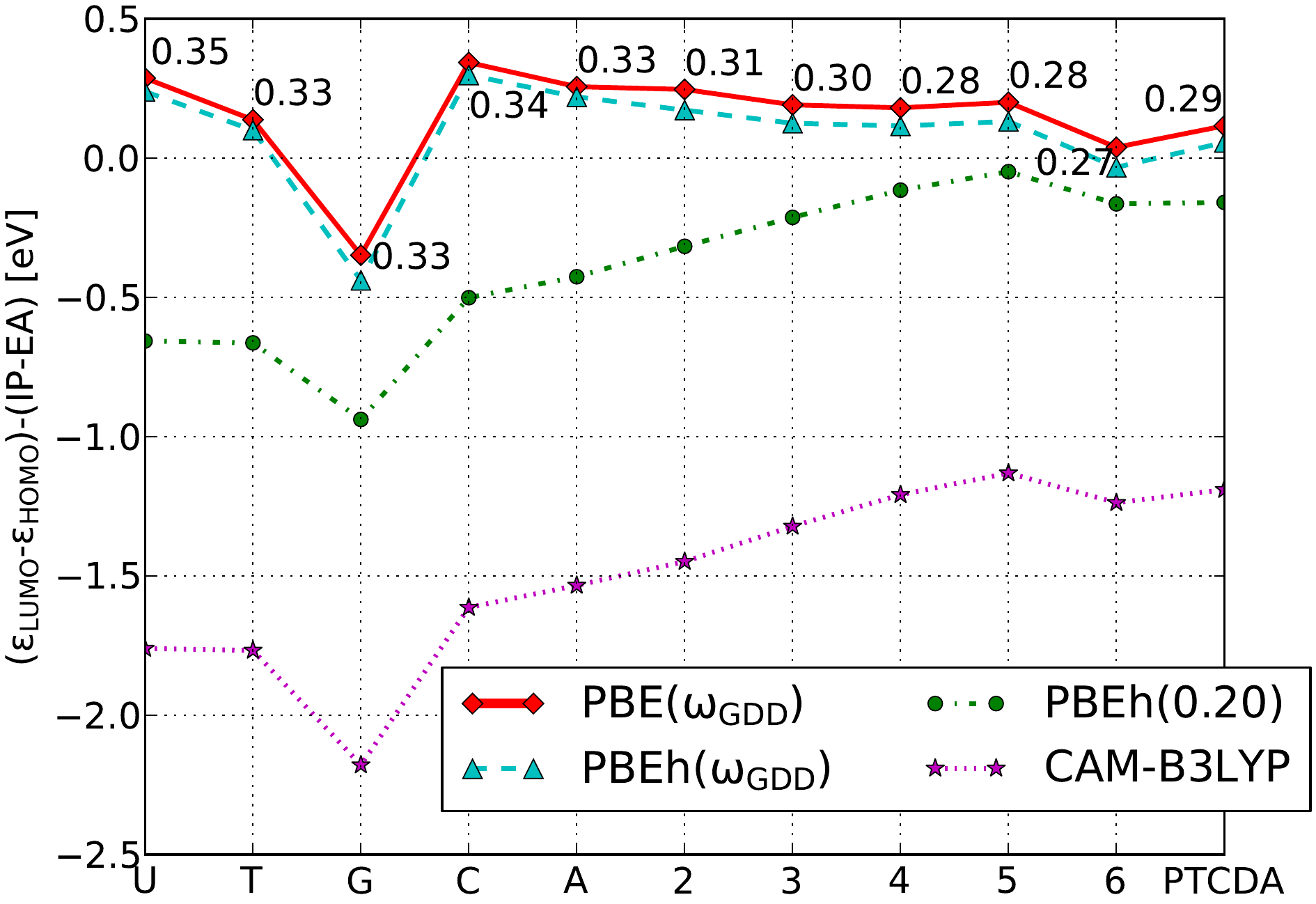}
\caption{Deviation of HL gaps from the theoretical fundamental gaps for DNA nucleobases\cite{bravaya2010electronic,epifanovsky2008electronically,faber2011first} (uracil, thymine, guanine, cytosine, adenine), linear acenes\cite{deleuze2003benchmark,hajgato2008benchmark} ($n=2,\ldots,6$), and PTCDA.\cite{blase2011first} The RS parameters for {\gddpbe} are given as annotations.}\label{fg-figure}
\end{figure}

To verify whether the GDD method provides physically meaningful LUMOs, the HL orbital gaps were calculated for nucleobases, a series of acenes, and perylene-3,4,9,10-tetracarboxylic dianhydride (PTCDA), see Fig.~\ref{fg-figure}. The anions of DNA nucleobases and anthracene have finite lifetimes. Their negative EAs are estimated by computing $\elumo$ in the finite basis set. The reference EAs were computed with the GW method\cite{faber2011first} (nucleobases) and coupled-cluster theory\cite{hajgato2008benchmark} (anthracene). In all cases, EAs have only a minor effect on the magnitude of the fundamental gap. {\gddpbe} and its short-range hybrid counterpart, {\gddpbeh}, yield HL gaps deviating from the reference theoretical fundamental gaps by merely $2.7\%$ and $2.1\%$ on average, respectively. The system-independent RS functionals, {\herbert} and CAM-B3LYP\cite{yanai2004new}, fail to achieve an uniform error distribution across all system sizes. The MAPDs are $4.4\%$ and $19.5\%$, respectively.


\begin{figure}
  \includegraphics[width=0.5\textwidth]{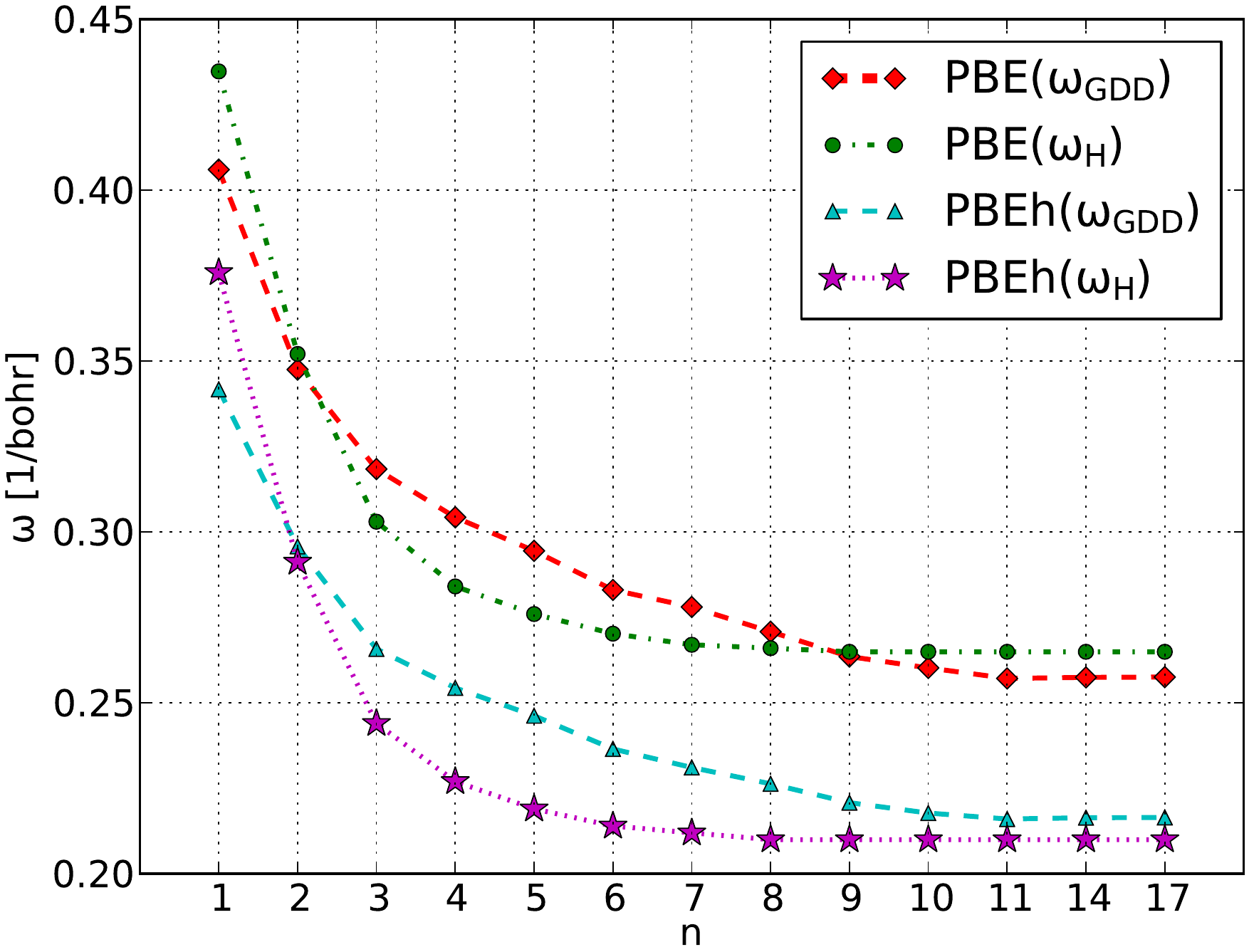}
  \caption{IP-optimized and GDD RS parameters for linear alkane chains, \ce{C_{2n}H_{4n+2}}. Calculations employed def2-TZVP basis.\cite{weigend2005balanced,schuchardt2007basis} The IP-optimized RS parameters and geometries were taken from Ref.~\citenum{koerzdoerfer2011long}.}\label{alkanes-figure}
\end{figure}

\begin{figure}
  \includegraphics[width=0.5\textwidth]{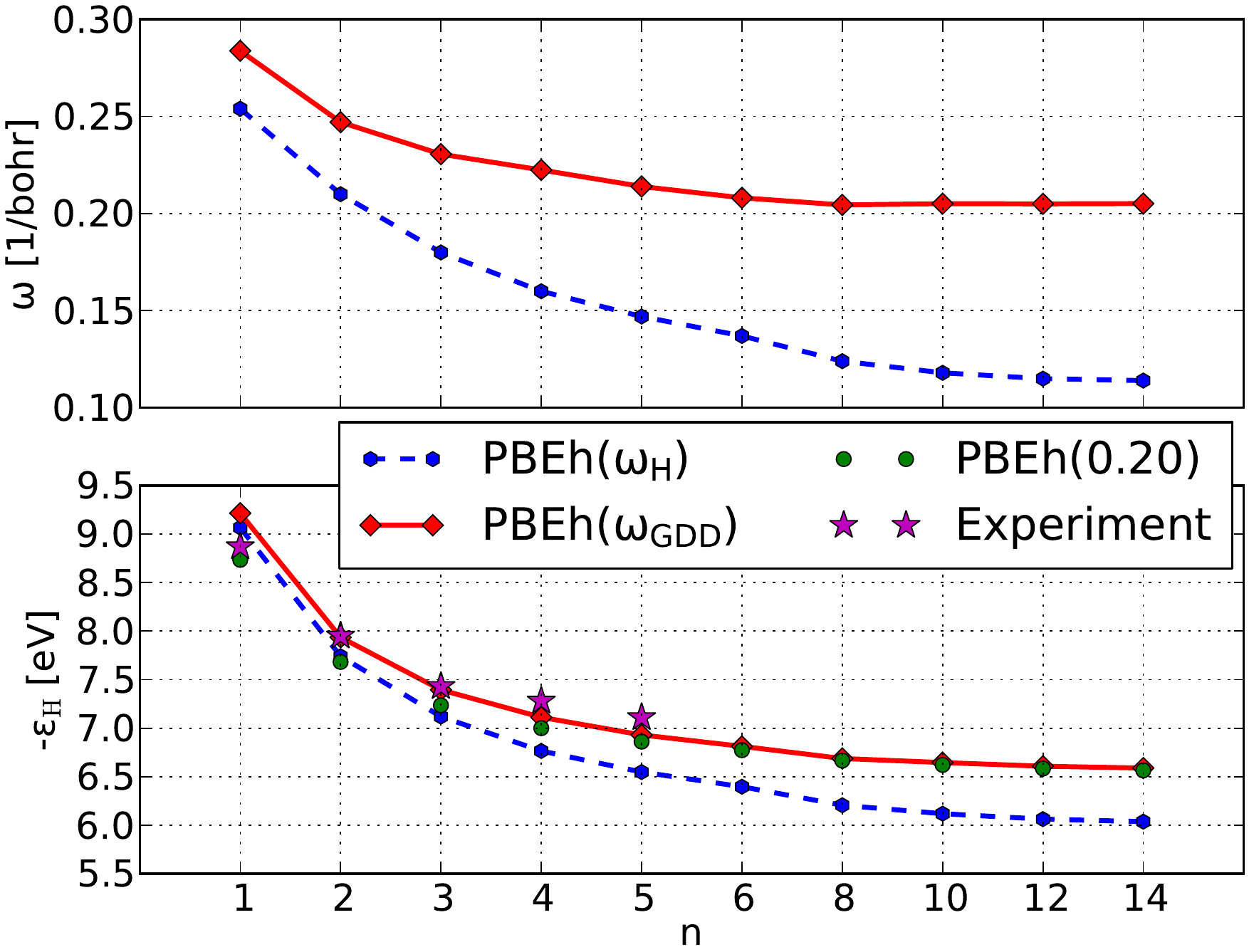}
  \caption{RS parameters (upper panel) and IPs from Koopmans' theorem (lower panel) for oligothiophene chains. Experimental IPs were taken from Ref.~\citenum{jones1990determination}. The values of $\oip$ were taken from Ref.~\citenum{koerzdoerfer2011long}.}\label{thiophenes-figure}
\end{figure}

 Both $\oip$ and $\ogdd$ decrease monotonically when computed for organic
 homopolymers of increasing length. In linear alkanes, $\ogdd$ closely
 follows $\oip$, see Fig.~\ref{alkanes-figure}. In contrast to the alkanes,
 we have observed a discrepancy between $\ogdd$ and $\oip$ in polymers
 with unsaturated bonds. For linear acenes, a decrease of $\ogdd$ with the chain length is less
 steep than that of $\oip$. The discrepancy grows with the chain length; for hexacene $\ogdd$ is
 equal to $0.27$~bohr$^{-1}$, whereas~\citet{koerzdoerfer2011long} report $\oip$ lower by $0.08$~bohr$^{-1}$.
 In this system, the HL orbital gap calculated with {\ippbe} is $0.49$~eV too low, whereas the {\gddpbe}
 functional is in excellent agreement with the reference CCSD(T) value,\cite{deleuze2003benchmark,hajgato2008benchmark}
 being only $0.04$~eV too high. In chains of oligothiophenes containing up to $14$ units,
 $\ogdd$ decreases less steeply than $\oip$ and saturates at a value nearly two times larger
 than $\oip$, see Fig.~\ref{thiophenes-figure}. For the shortest chains ($n=2,\ldots,5$),
 the IPs calculated with {\gddpbeh} agree with the experimental data\cite{jones1990determination} to
 within $0.2$~eV. {\ippbeh} increasingly deviates from the experimental IPs as the chain lengthens.
 For a 14-unit oligothiophene, the IPs obtained with {\gddpbeh} and {\herbert} are closer to a rough
 estimate of the IP in an infinite oligothiophene chain\cite{jones1990determination} than the IP
 obtained with {\ippbeh}.

\begin{figure}
  \includegraphics[width=0.5\textwidth]{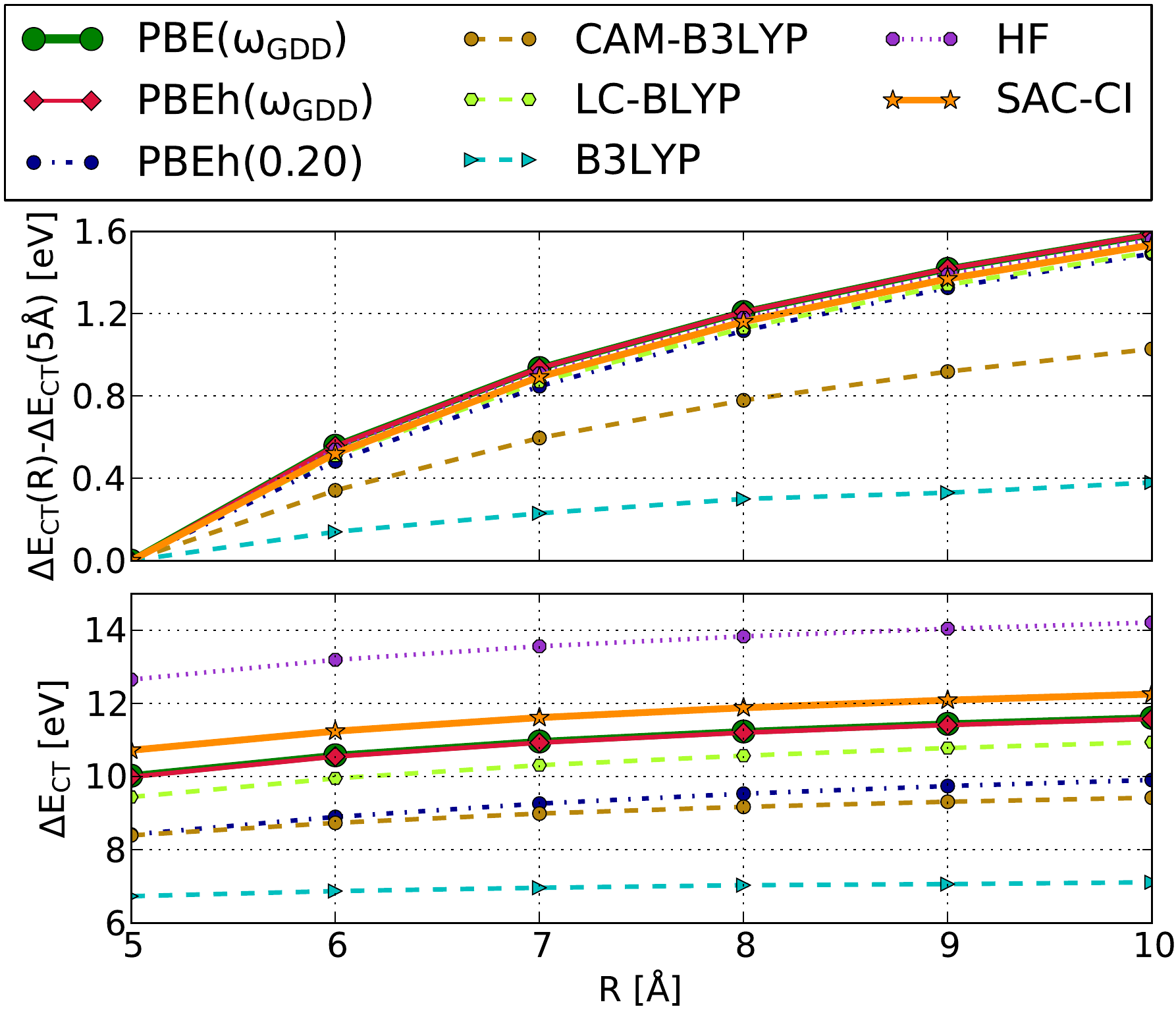}
\caption{Lowest triplet CT excitation of \ce{C2H4{\cdots}C2F4} for several intermolecular distances ($R$). The reference SAC-CI\cite{nakatsuji1978cluster} employed cc-pVTZ basis set. The energies for HF, B3LYP, and LC-BLYP are taken from Ref.~\cite{tawada2004long}.} \label{c2h4_c2f4}
\end{figure}

Frontier orbital energies are related to the lowest CT excitation energy in adiabatic time-dependent DFT through Mulliken's rule,\cite{dreuw2003long}
\begin{equation}
\Delta E_\mathrm{CT} \rightarrow \mathrm{IP}-\mathrm{EA} -\frac{1}{R} = \elumo - \ehomo - \frac{1}{R}, \label{mulliken}
\end{equation}
where $R$ denotes a donor---acceptor separation. Because the GDD method was demonstrated to
 align orbital energies with IPs, we expect that it can accurately describe CT excitations.
 Indeed, data in Table~\ref{tcne-table} show that the GDD method recovers the gas-phase experimental
 CT excitation energies of aromatic donor (Ar)-tetracyanoethylene (TCNE) noncovalent complexes\cite{hanazaki1972vapor}
 at the accuracy similar to the GW-Bethe-Salpeter method\cite{blase2011charge} and
 the optimally-tuned RS BNL functional.\cite{stein2009reliable} A system-independent RS
 functional, {\herbert}, is characterized by a five times larger MAD for these systems.
 Fig.~\ref{c2h4_c2f4} shows the distance dependence of CT excitations in \ce{C2H4\bond{...}C2F4}
 complex. The SAC-CI\cite{nakatsuji1978cluster} energies used as the reference were obtained
 with larger basis set and are considerably lower than the values previously reported by \citet{tawada2004long}
 All the RS functionals (LC-BLYP, CAM-B3LYP, {\gddpbe}, and {\gddpbeh}) as well as HF yield accurate energy differences
 relative to the values at $5.0$~\AA. However, the absolute values of excitation energies differ
 among the methods by several electronvolts, as seen in the lower panel of Fig.~\ref{c2h4_c2f4}.
 The constant shift by which the methods differ can be attributed to different values of the
 HL gap in \eqn{mulliken}. Of all the tested methods, the GDD-based RS functionals are
 the closest to the benchmark curve.

\begin{table}[H]
\caption{Energies and oscillator strengths (in parentheses) of CT transitions in gas phase Ar-TCNE complexes. All energies are given in electronvolts. 
} \label{tcne-table}
\begin{tabular}{llllll}
\hline \hline
Ar                  & \gddpbe & \gddpbeh      & BNL\cite{stein2009reliable} &  GW\cite{blase2011charge} & exp.\cite{hanazaki1972vapor} \\
\hline
toluene             &  3.33(0.03) &  3.23(0.03) & 3.4  & 3.27 & 3.36(0.03) \\
o-xylene            &  3.36(0.03) &  3.26(0.03) & 3.0  & 2.89 & 3.15(0.05) \\
benzene             &  3.71(0.02) &  3.60(0.02) & 3.8  & 3.58 & 3.59(0.02) \\
\ce{C10H8}          &  2.63(0.00) &  2.52(0.00) & 2.7  & 2.55 & 2.60(0.01) \\
\hline
MAD                 & 0.10       & 0.08       & 0.12 & 0.10 &   \\
\hline \hline
\end{tabular}
\end{table}

\begin{figure}
  \includegraphics[width=0.5\textwidth]{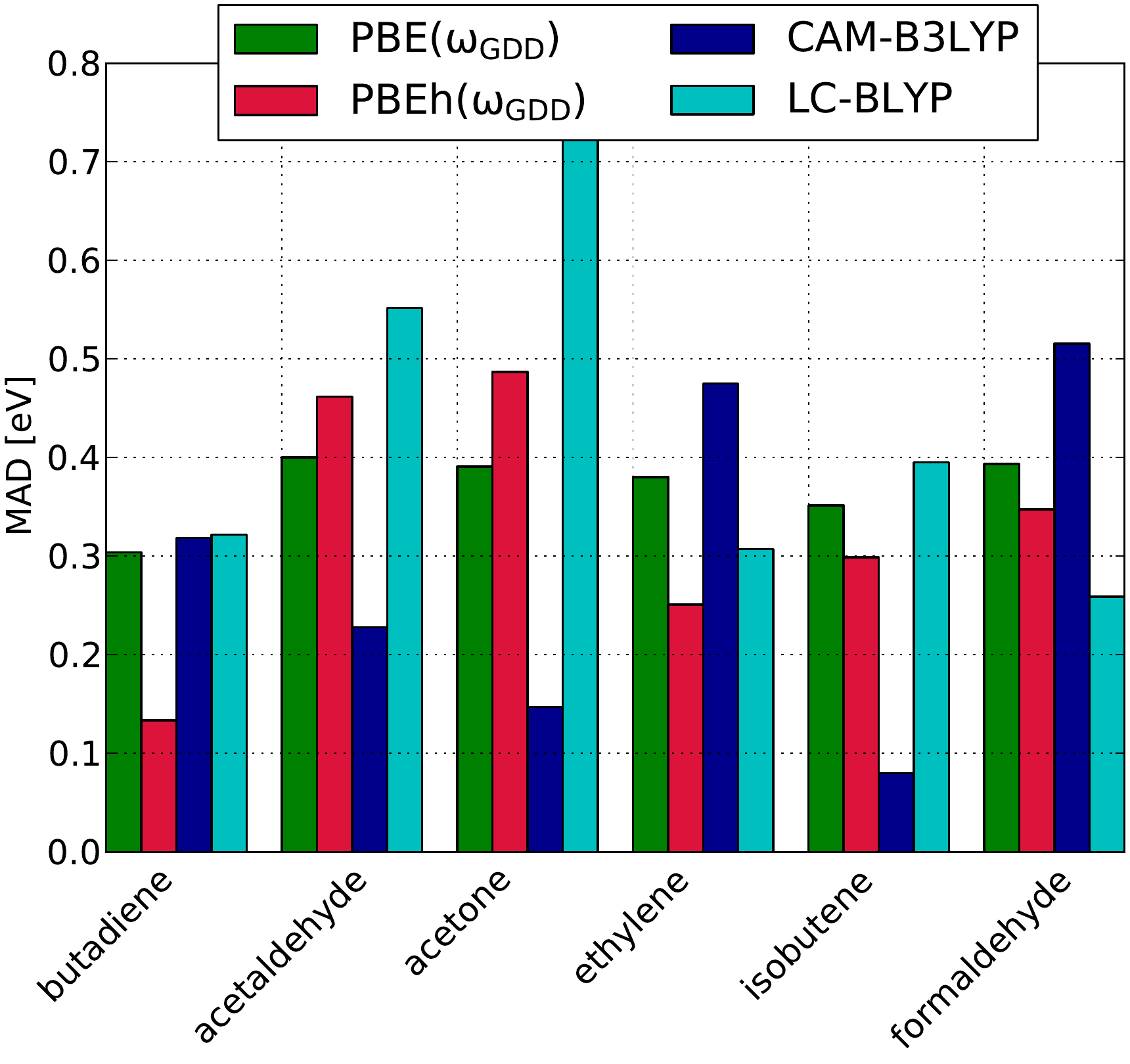}
  \caption{MADs between TDDFT and experimental energies of Rydberg transitions. All Rydberg transitions compiled in the database by~\citet{caricato2010electronic} were used.} \label{rydberg-mads}
\end{figure}

One may expect that a method for which $-\ehomo \approx \mathrm{IP}$ should at the same time yield accurate energies of Rydberg transitions. To test this hypothesis, we gauged the performance of GDD-based functionals in a set of small molecules for which experimental energies of Rydberg transitions are available: butadiene, acetaldehyde, acetone, ethylene, isobutene, and formaldehyde. The respective MADs are presented in Fig.~\ref{rydberg-mads}. The GDD computations employed 6-311(3+,3+)G** basis set.\cite{caricato2010oscillator} The transition assignments, experimental energies, and the results for LC-BLYP and CAM-B3LYP were taken from Ref.~\citenumwrap{caricato2010electronic}. We have found that, statistically, the GDD method does not lead to such a pronounced accuracy gain over the other methods as in the case of CT transitions. Even in the case of ethylene, where HOMO energies are within 0.1~eV of the experimental $-\text{IP}$, the Rydberg transitions are systematically overestimated by about 0.3~eV by both {\gddpbe} and {\gddpbeh}, see Table~\ref{rydbergs-ethylene}. We conclude that, in the case of Rydberg transitions, an approximate alignment of $\ehomo$ with -IP prevents huge outlying absolute errors, but does not guarantee reduction of MADs.

\begin{table}
  \caption{Rydberg excitations in ethylene. All energies are given in electronvolts. The results for CAM-B3LYP, LC-BLYP, and EOM-CCSD are taken from Ref.~\citenum{caricato2010electronic}. The experimental vertical IP is taken from Ref.~\citenum{linstrom2013nist}.}\label{rydbergs-ethylene}
  \begin{tabular}{lllllll}
  \hline \hline
  &\gddpbe&\gddpbeh&CAM-B3LYP&LC-BLYP&EOM-CCSD&exp. \\
  \hline
  $1B_{3u}$&7.61&7.49&6.89&7.41&7.28&7.11 \\
  $1B_{1g}$&8.20&8.10&7.48&8.04&7.93&7.80 \\
  $1B_{2g}$&8.37&8.23&7.54&8.21&7.96&7.90 \\
  $2A_g$  &8.59&8.47&7.87&8.52&8.31&8.28 \\
  $2B_{3u}$&9.09&8.96&8.26&8.99&8.80&8.62 \\
  $3B_{3u}$&9.33&9.20&8.43&9.27&9.06&8.90 \\
  $4B_{3u}$&9.43&9.30&8.55&9.43&9.16&9.08 \\
  $3B_{1g}$&9.54&9.40&8.53&9.53&9.28&9.20 \\
  $2B_{1u}$&9.48&9.34&8.58&9.50&9.28&9.33 \\
  $5B_{3u}$&9.89&9.76&8.85&9.90&9.62&9.51 \\
  \hline
  MAD       & 0.38&0.25 &0.48 &0.31& 0.11 & \\
  $-\ehomo$ & 10.78 & 10.63 &     &    & &10.68 \\
  \hline \hline
  \end{tabular}
\end{table}


Similarly to other approaches employing system-specific RS parameter, the GDD-based functionals are not size-consistent. The size consistency can be retained by an RS functional only if $\omega$ is fixed for all systems, as in the {\herbert} functional. However, fixed-$\omega$ functionals violate other constraints of the exact functional: Koopmans' condition, the straight-line behavior of $E(N)$ between integers, and scaling to the high-density limit.\cite{krukau2008hybrid} In contrast, optimally-tuned as well as GDD-based functionals satisfy these conditions at least approximately. The choice between a fixed-$\omega$ and system-specific functional is dictated by the constraints which are important for a particular application. For example, fixed-$\omega$ functionals are preferred for computing enthalpies of formation,\cite{karolewski2013using} for which the size-consistency is crucial. On the other hand, there is numerical evidence that optimally-tuned functionals can be successfully applied to compute non-covalent binding energies,\cite{agrawal2013pair} provided that the optimal RS parameter is determined for the dimer and the same value is also used for the monomers. Also CT excitations in non-covalent systems can be excellently described by the system-specific methods, despite their lack of size-consistency, see the results for \ce{Ar\bond{...}TCNE} complexes in Table~\ref{tcne-table} herein and in Ref.~\citenum{stein2009reliable}. In case of non-covalent interactions, it is even possible to eliminate the lack of size-consistency of optimally-tuned or GDD-based functionals altogether by applying different values of $\omega$ for each of the interacting species within SAPT(DFT) framework\cite{misquitta2005intermolecular} or subsystem approaches.\cite{rajchel2010density}

\section{Conclusions}
To summarize, we devised a model of the electron---exchange hole interaction in the outer density region that leads to straightforward predictions of the interelectron distance at which the nonlocal exchange should supersede the local DFA. The model spawns a class of approximations based on existing RS functionals. Two of such approximations, {\gddpbe} and {\gddpbeh}, have been shown to align orbital energies with IPs and accurately describe CT excitations. The GDD method improves upon the existing approaches to the range separation in the following ways. \begin{inparaenum}[(i)] \item It is the first method capable of aligning single particle energies with IPs without the need for iterative calculation of ionized species. \item Unlike the functionals with a fixed RS parameter, such as CAM-B3LYP and {\herbert}, its performance is not affected by the system size. \item {$\ogdd$} depends on the global electron density and is constant in space, thus the computational cost is lower than in the local RS model of~\citetwrap{krukau2008hybrid}. \item It offers superior quality of the frontier orbital energies in long unsaturated polymers such as oligothiophenes and acenes, for which the IP-tuning procedure predicts spuriously low values of the $\oip$ parameter.\end{inparaenum}

\section{Acknowledgments}
This work was supported by the Polish Ministry of Science and Higher Education under Grant No.~NN204248440, and by the National Science Foundation under Grant No.~CHE-1152474. ŁR acknowledges the support of the Alexander von Humboldt Foundation.

\bibliography{biblio}
\end{document}